\RequirePackage{fix-cm}
\documentclass{svjour3}                     
\smartqed  
\usepackage{graphicx}
\usepackage{hyperref}
\newcommand\etal {{\it et al.}}

\newcommand\al{\alpha}

\newcommand\ga{\gamma}
\newcommand\de{\delta}

\newcommand\ka{\kappa}
\newcommand\la{\lambda}

\newcommand\si{\sigma}

\newcommand\ps{\psi}

\newcommand\Ga{\Gamma}

\newcommand\Si{\Sigma}

\newcommand\mn{{\mu\nu}}

\newcommand\fr[2]{{{#1} \over {#2}}}
\newcommand\half{{\textstyle{1\over 2}}}

\newcommand\fracn[2]{{\textstyle{{#1}\over {#2}}}}

\newcommand\lsim{\mathrel{\rlap{\lower4pt\hbox{\hskip1pt$\sim$}}
    \raise1pt\hbox{$<$}}}
\newcommand\gsim{\mathrel{\rlap{\lower4pt\hbox{\hskip1pt$\sim$}}
    \raise1pt\hbox{$>$}}}
\newcommand\sqr[2]{{\vcenter{\vbox{\hrule height.#2pt
         \hbox{\vrule width.#2pt height#1pt \kern#1pt
         \vrule width.#2pt}
         \hrule height.#2pt}}}}

\newcommand\thpr{{these proceedings}}

\newcommand\pt[1]{\phantom{#1}}
\newcommand\ol[1]{\overline{#1}}

\newcommand\vb[2]{e_{#1}^{{\pt{#1}}#2}}
\newcommand\ivb[2]{e^{#1}_{{\pt{#1}}#2}}
\newcommand\uvb[2]{e^{#1#2}}

\newcommand\ab{\overline{a}{}}

\newcommand\cb{\overline{c}{}}

\newcommand\sbar{\overline{s}{}}

\newcommand\mt{m^{\rm T}}
\newcommand\ms{m^{\rm S}}

\newcommand\afb{(\ab_{\rm{eff}})}

\newcommand\afbx[1]{(\ab^{#1}_{\rm{eff}})}
\newcommand\cbx[1]{(\cb^{#1})}

\newcommand\afbe{\afbx{e}}

\newcommand\cbw{\cbx{w}}
\newcommand\afbw{\afbx{w}}

\newcommand\lrpartial{\raise 1pt\hbox{$\stackrel\leftrightarrow\partial$}}

\newcommand\lrDmu{\stackrel{\leftrightarrow}{D_\mu}}

\newcommand\atext{$a_\mu$}
\newcommand\btext{$b_\mu$}
\newcommand\ctext{$c_{\mu\nu}$}
\newcommand\dtext{$d_{\mu\nu}$}
\newcommand\etext{$e_\mu$}
\newcommand\ftext{$f_\mu$}
\newcommand\gtext{$g_{\la\mu\nu}$}
\newcommand\Htext{$H_{\mu\nu}$}

\newcommand\G{G_N}

\newcommand{\beq}{\begin{equation}}
\newcommand{\eeq}{\end{equation}}
\newcommand{\bea}{\begin{eqnarray}}
\newcommand{\eea}{\end{eqnarray}}
\newcommand{\bit}{\begin{itemize}}
\newcommand{\eit}{\end{itemize}}

\newcommand\pno[1]{PNO(#1)}

\journalname{Hyperfine Interactions (2012) 213:137-146}
\begin{document}

\title{Antimatter, the SME, and Gravity
}


\author{Jay D.\ Tasson 
}


\institute{Jay D.\ Tasson \at
              Department of Physics\\
	      Whitman College\\
	      Walla Walla, WA 99363, USA\\
              \email{jtasson@carleton.edu}           
}

\date{Published: 12 September, 2012.  The original publication is available
at \href{http://link.springer.com/article/10.1007/s10751-012-0642-3}{www.springerlink.com}.}


\maketitle

\begin{abstract}
A general field-theoretic framework
for the analysis of CPT and Lorentz violation
is provided by the Standard-Model Extension (SME).
This work discusses
a number of SME-based proposals for tests of CPT and Lorentz symmetry,
including antihydrogen spectroscopy
and antimatter gravity tests.
\keywords{Antimatter \and Lorentz violation \and Gravity}
\PACS{11.30.Cp \and 04.80.Cc \and 11.30.Er}
\end{abstract}

\section{Introduction}
\label{intro}

The Standard Model of particle physics
along with General Relativity 
provide an excellent description of known physics.
As a foundational principle of each
of our best theories,
Lorentz symmetry,
along with the associated CPT symmetry \cite{cpt},
should be well tested experimentally.
It is also likely that the Standard Model
and General Relativity 
are limits of a more fundamental theory
that provides consistent predictions all the way to the Planck scale.
A technically feasible means
of searching for potential suppressed signals from 
a complete theory at the Planck scale
is provided by tests of CPT and Lorentz symmetry \cite{ksp}.
A comprehensive test framework
for searching for such potential signals
across all areas of known physics
is provided by the SME \cite{ck,akgrav}.

\section{Theory}
\label{theory}

The basic theory relevant for the discussion to follow
is the QED extension limit of the gravitationally coupled SME \cite{akgrav}.
Schematically,
the action for the theory can be written
\beq
S = S_\ps + S_A + S_{\rm gravity}.
\label{action}
\eeq
The first term here is the gravitationally coupled fermion sector,
the second is the photon sector,
and the final term is the pure-gravity sector.
Each of the above terms
consists of known physics,
followed in general by all Lorentz-violating terms.
Here we consider what is known as the minimal SME,
which is the restriction to operators of dimension 3 and 4.

The minimal fermion-sector action can be written
\beq
S_\ps = 
\int d^4 x (\half i e \ivb \mu a \ol \ps \Ga^a \lrDmu \ps 
- e \ol \ps M \ps).
\label{fermion}
\eeq
where
\bea
\Ga^a
&\equiv & 
\ga^a - c_{\mu\nu} \uvb \nu a \ivb \mu b \ga^b
- d_{\mu\nu} \uvb \nu a \ivb \mu b \ga_5 \ga^b
\nonumber\\
&&
- e_\mu \uvb \mu a 
- i f_\mu \uvb \mu a \ga_5 
- \half g_{\la\mu\nu} \uvb \nu a \ivb \la b \ivb \mu c \si^{bc}, \\
M
&\equiv &
m + a_\mu \ivb \mu a \ga^a 
+ b_\mu \ivb \mu a \ga_5 \ga^a 
+ \half H_{\mu\nu} \ivb \mu a \ivb \nu b \si^{ab},
\label{mdef}
\eea
and \atext, \btext, \ctext, \dtext, \etext, \ftext, \gtext, \Htext\ 
are coefficient fields for Lorentz violation.
Couplings to gravity
occur here via the vierbein $\vb \mu a$
and contributions to the covariant derivative.
The form of the Minkowski-spacetime fermion-sector action
can be recovered by taking the limit $\vb \mu a \rightarrow \de^a_\mu$.

The action $S_A$
provides the photon sector.
It contains Maxwell electrodynamics
followed in the minimal case by Lorentz-violating terms
of dimension 3 and 4,
though operators of arbitrary dimension
have now been classified,
and numerous new experimental proposals 
associated with them have been made \cite{nonmin}.
Though generically of considerable interest,
the explicit form of $S_A$ is omitted here
since it is not directly relevant for the discussion to follow.

The minimal contributions to the pure-gravity sector
take the form
\beq
S = \fr {1}{2\ka} \int d^4x e (R - u R 
+s^\mn R^T_\mn + t^{\ka\la\mu\nu} C_{\ka\la\mu\nu}),
\label{grav}
\eeq
where $R^T_\mn$ is the traceless Ricci tensor,
and $C_{\ka\la\mu\nu}$ is the Weyl tensor.
The standard Einstein-Hilbert action
is provided by the first term.
The coefficient field $s^\mn$ in the third term
is responsible for
the relevant Lorentz-violating signals 
in the post-Newtonian analysis \cite{lvpn}.
The fourth term provides no contributions
in the post-Newtonian analysis,
while the second term is not Lorentz violating.

The Minkowski-spacetime limit
of action \ref{action}
forms the basis of the 
discussion of nongravitational tests in Sec.\ \ref{ngt}.
Here
the pure-gravity sector action \ref{grav}
is irrelevant and gravitational couplings
in the fermion-sector action \ref{fermion}
can be neglected.
Relevant theoretical techniques for the analysis of a variety of experiments
have been developed.
The relativistic quantum mechanics 
associated with action \ref{fermion}
has been developed after addressing a technical point about how to achieve
a hermitian Hamiltonian \cite{bkr}.
The corresponding nonrelativistic Hamiltonian can then be achieved
via a standard Foldy-Wouthuysen procedure \cite{kl}.
The associated classical Lagrangian has also been obtained \cite{kr}.
Applications of these Minkowski-spacetime results
to antimatter are considered in Sec.\ \ref{ngt}.

Gravitational couplings are the focus of Sec.\ \ref{gt}.
The necessary results for the analysis of tests
based on Lorentz violation in the gravity sector 
are developed in Ref.\ \cite{lvpn}.
Though more general geometries may admit explicit breaking \cite{akfin},
in the context of gravitational theory based on Riemannian geometry,
Lorentz violation must arise spontaneously \cite{akgrav}.
A key theoretical challenge 
addressed in Ref.\ \cite{lvpn} was the role of the fluctuations
in the coefficient fields \cite{rbak}.
Upon meeting this challenge,
the metric and the equations of motion relevant for post-Newtonian tests
were obtained,
and numerous detailed experimental proposals for investigating
the coefficient (vacuum value) $\sbar^\mn$ 
associated with the coefficient field $s^\mn$
were made.
These experimental proposals
and some associated experimental results
are discussed further in Sec.\ \ref{grr}.

Many of the methods and results of the nongravitational analysis
of the fermion sector discussed above
are extended and combined with the gravitational sector results
in addressing the gravitational couplings in the fermion sector,
which are the subject of Ref.\ \cite{lvgap}.
This includes the implications of the fluctuations in the coefficient fields
that must be addressed in a gravitational analysis.
Some additional theoretical issues are also addressed
before specializing to the case of spin-independent couplings
for the consideration of experiments.
This limit consists of the \atext, \ctext,\ and \etext\
coefficient fields,
with the associated vacuum values being denoted $\afb_\mu$,
for the countershaded combination \cite{akjt}
of the vacuum values associated with the \atext\ and \etext\ fields,
and $\cb_\mn$ for the vacuum value associated with \ctext. 
These vacuum values correspond to the coefficients for Lorentz violation
discussed in Minkowski spacetime.
Many of the relevant experiments
are the same as those that are relevant 
for investigating the pure-gravity sector
coefficient $\sbar^\mn$,
but the associated signals can be quite different.
The experimental implications are summarized in Sec.\ \ref{grr},
while specific implications for antimatter are discussed in Sec.\ \ref{gat}.

\section{Nongravitational Tests with Antimatter}
\label{ngt}

Numerous experimental investigations have been performed
in the context of the SME \cite{tables}.
One effect of the Lorentz-violating terms in the SME
that has been exploited experimentally
is modifications to the energy levels of atoms \cite{clock,hhbar}.
Lorentz violation can then be detected though a comparisons
of levels.
As the Earth rotates on its axis
and revolves around the Sun,
the orientation and boost of an experiment in the lab change.
In the present context of studies of Lorentz violation,
this implies an annual and sidereal variation of energy levels.
Since the effects on various levels may differ,
Lorentz violation can be investigated
by searching for relative changes in the levels with time  \cite{clock}.

A subset of the Lorentz-violating terms in the SME
are also CPT violating.
Thus the comparison of the spectrum of hydrogen
with that of antihydrogen
would also provide sensitivity \cite{hhbar},
and an experiment aiming to obtain sensitivity to SME coefficients
via a measurement of the hyperfine Zeeman line
is presently in preparation \cite{bjhyp}.
In some cases,
Lorentz and CPT violating effects
that are difficult to detect in matter experiments
may be much more readily seen
in comparisons of matter and antimatter.

A simple toy model 
called the isotropic `invisible' model (IIM) \cite{akifc,lvgap}
was constructed
to illustrate a scenario in which Lorentz violation
would produce a significantly larger effect in 
tests involving antimatter
than in tests with only matter.
To construct the model,
an isotropic (`fried-chicken') limit of the SME
is chosen.
In any give inertial frame $O$,
a subset of Lorentz-violating operators
in the SME preserve rotational symmetry.
Setting the coefficients of the other operators to zero
produces isotropic Lorentz violation
in that frame.
Another frame $O^\prime$ boosted with respect to $O$
will not have rotation invariance,
that is,
the effects of Lorentz violation
are isotropic in $O$ but not in $O^\prime$.  

The IIM assumes
the only nonzero coefficients for Lorentz violation
in the CMB frame
are $(b^p)_{T^\prime}$ and isotropic $(d^p)_{\Xi^\prime\Xi^\prime}$.
The nonzero coefficients are chosen to obey the condition
$(b^p)_{T^\prime} = k m^p (d^p)_{T^\prime T^\prime}$,
for a suitable choice of constant $k$.
Nonzero coefficients
$(b^p)_J$ and $(d^p)_{JT}$
are then generated in the Sun-centered frame.
In terrestrial experiments with hydrogen,
the dominant signals
appear in the hyperfine structure
and are due to the combination
$(b^p)_J - m^p (d^p)_{JT}$,
which vanishes for suitable $k$.
Therefore in the context of the IIM,
these experiments can only detect
effects suppressed by at least one power
of the boost of the Earth around the Sun,
which is about $10^{-4}$
and requires an experiment sensitive to annual variation.
The dominant effects in experiments with antihydrogen,
however,
involve the combination
$(b^p)_J + m^p (d^p)_{JT}$.
Thus unsuppressed signals would occur in the hyperfine structure
of antihydrogen.
The IIM is thus a field-theoretic toy model
in which the effects of CPT and Lorentz violation
would be at least 10,000 times greater in antihydrogen
than those in hydrogen
or other nonrelativistic neutral matter.

In addition to shifting the atomic energy levels,
the energy levels of trapped particles
are shifted.
Thus the comparison of the energy levels for trapped particles
with those of trapped antiparticles
also leads to sensitivities to CPT violation in the SME \cite{bkr}.
Results have been obtained
for coefficients for Lorentz violation associated with the 
electron \cite{etrap}
and proton \cite{ggtrap}
via this method
and ongoing experiments provide the possibility of further results \cite{wq}.

\section{Gravitational Tests}
\label{gt}

\subsection{Overview of Experiments}
\label{grr}

A wide variety of gravitational experiments
can achieve sensitivity to 
coefficients $\afb_\mu$, $\cb_\mn$, and $\sbar_\mn$.
One diverse class of such experiments
are those that can be performed in Earth-based laboratories.
Proposed or performed tests of this type
include gravimeter experiments,
tests of the universality of free fall,
and experiments with devices 
traditionally used for short-range gravity tests.

The key point found in the 
analysis of Refs.\ \cite{lvpn} and \cite{lvgap}
is that the gravitational force acquires tiny corrections
both along and perpendicular to the usual free-fall trajectory
near the surface of the Earth
in the presence of coefficients $\afb_\mu$, $\cb_\mn$, and $\sbar_\mn$.
Modifications to the effective inertial mass of a test body
that are direction dependent
are also generated by coefficients $\afb_\mu$ and $\cb_\mn$,
which results in a nontrivial relation between force and acceleration.
The corrections considered here are time dependent with variations
at the annual and sidereal frequencies.
Additionally,
the corrections introduced by
$\afb_\mu$ and $\cb_\mn$
are particle-species dependent.

The above discussion leads naturally
to a 4-category classification
of 
laboratory tests using Earth as a source.
Free-fall gravimeter tests
and force-comparison gravimeter tests
make use of the characteristic time dependencies introduced,
by monitoring the gravitation acceleration or force over time
respectively.
The relative acceleration of,
or relative force on,
a pair of test bodies
can also be considered.
Experiments that measure these quantities
constitute free-fall and force-comparison Weak Equivalence Principle 
(WEP) tests
respectively.
Devices presently used or considered for the above types of tests
include 
falling corner cubes \cite{fc},
atom interferometers \cite{ai,aigrav,mh},
gravimeters based on superconducting levitation \cite{fcgrav},
laboratory masses tossers \cite{tossed},
balloon drop equipments \cite{balloon},
drop towers \cite{bremen},
sounding rockets \cite{srpoem},
and
torsion pendula \cite{tpwep}.
Specific predictions
and estimated sensitivities for the above tests,
including a frequency decomposition of the relevant signal
to which experimental data could be fit,
are provided by Refs.\ \cite{lvpn} and \cite{lvgap}.
A test of this type has already been performed
using an atom-interferometer \cite{aigrav}.
Note that the effective WEP violation considered here
is qualitatively different from that sought in WEP tests to date
due to the characteristic time dependence.

Though less sensitive at present
to the range-independent SME effects presently under discussion,
systems in which both the source mass 
and the test mass are contained within the lab,
such as those
devices 
traditionally used as tests of gravity at short range,
can also be considered \cite{db,cspeake}.
Space-based versions of the WEP tests discussed above
have also been proposed \cite{space},
with the potential for
significant sensitivity improvements.
Lorentz-violating signals in such tests 
would again have characteristic time dependence
distinguishing them from other sources of WEP violation.
An SME analysis has also been performed \cite{lvpn}
for the space-based Gravity Probe B experiment \cite{gpb}.

Experimentally challenging versions
of the tests highlighted above
performed with  antimatter, charged particles,
and second- and third-generation particles
can yield sensitivities to Lorentz and CPT violation 
that are otherwise difficult or impossible to achieve.
Gravitational experiments with antihydrogen
\cite{gblmstp,tjpage,jwth,bouncebar,weax,aegis},
charged-particle interferometry \cite{chargeai},
ballistic tests with charged particles \cite{charge},
and signals in muonium free fall \cite{muon}
are considered in Ref.\ \cite{lvgap}.
Antihydrogen tests are discussed in more detail
in Sec.\ \ref{gat} of this work.

The observations of orbits also provides a means of 
searching for Lorentz-violating effects in gravitational physics.
References \cite{lvpn} and \cite{lvgap}
consider tests that search for such effects
in lunar laser ranging \cite{llrsme},
perihelion precession measurements \cite{peri},
and binary-pulsar observations \cite{pulsar}.

The interaction of photons with gravity
as well as effects on the clocks typically associated
with such tests
provides a final class of investigations.
Signals for Lorentz violation arising 
in measurements of the time delay,
gravitational Doppler shift,
and gravitational redshift,
along with
comparisons of the behaviors of photons and massive bodies
have been considered for both the pure-gravity sector \cite{qbdoppler}
and the matter sector \cite{lvgap}.
Null redshift tests
are also relevant in the context of the matter sector \cite{lvgap}.
The implications for a variety of experiments
and space missions \cite{photon} are also discussed
in Refs.\ \cite{qbdoppler}
and \cite{lvgap}.
Sensitivities
to $\afb_\mu$ and $\cb_\mn$ coefficients have been achieved
based on this work via an analysis of a variety of clocks \cite{mh}.
Note that these results
and proposals are in addition to the SME-based clock tests
being performed on the ground 
and proposed for space \cite{clock}.

\subsection{Antimatter Tests}
\label{gat}

A number of methods for measuring the gravitational acceleration
of antihydrogen have been suggested
as ways of probing antimatter-gravity interactions.
Some such ideas include tests using
trapped antihydrogen 
\cite{gblmstp},
antihydrogen interferometry 
\cite{tjpage},
antihydrogen free fall from an antiion trap
\cite{jwth},
bouncing antihydrogen
\cite{bouncebar},
and tests in space
\cite{weax}.
The Antimatter Experiment: Gravity, Interferometry, Spectroscopy
(AEGIS)
\cite{aegis},
with an initial sensitivity goal of 1\%
to the gravitational acceleration of antihydrogen,
provides an example of the initial sensitivity
likely to be achieved.

In the context of gravitational couplings
in the fermion sector of the SME,
antimatter-gravity experiments could obtain special sensitivities
to the coefficients $\afbw_\mu$ and $\cbw_\mn$.
The idea is that a CPT transformation
reverses the sign of $\afbw_\mu$ 
but not the sign of $\cbw_\mn$.
As a result,
experiments with antihydrogen 
could in principle observe novel behaviors
and could place cleaner constrains
on certain combinations of SME coefficients
than can be obtained with matter.
The theoretical treatment 
of relevant antimatter experiments
is then the same as the treatment of the matter experiments
discussed above
with the only exception being the change in sign
of $\afbw_\mu$ relative its matter counterpart
throughout the analysis.

In addition to providing a framework 
for the analysis of antimatter gravity experiments,
the general field-theoretic approach of the SME
elucidates aspects of the numerous attempts
to place indirect limits on the possibility of unconventional 
antimatter-gravity interactions
that appear in the literature \cite{mntg}.
The discussion proceeds most efficient when presented in the context
of an explicit toy-model limit
of the SME,
the isotropic `parachute' model (IPM),
which is similar in design to the IIM discussed above.

The IPM is constructed
by restricting the classical nonrelativistic Lagrange density of the SME  
in the Sun-centered frame $S$,
to the limit in which only coefficients 
$\afbw_T$ and isotropic $\cbw_{\Si\Xi}$ are nonzero.
In this limit,
the effective classical Lagrangian 
for a test particle T moving in the gravitational field
of a source S
can be written in the suggestive form
\beq
L_{\rm IPM} = \half \mt_i v^2 + \fr{\G \mt_g \ms_g}{r},
\eeq
where $\mt_i$ is the effective inertial mass of T,
and $\mt_g$ and $\ms_g$ 
are the effective gravitational masses 
of T and S, respectively.
The effective masses are defined in terms of 
the coefficients $\afbw_T$, $\cbw_{TT}$ for Lorentz violation
and the conventional Lorentz invariant body masses $m^{\rm B}$
as follows:
\bea
\nonumber
m^{\rm B}_i &=& 
m^{\rm B} + \sum_w \fracn53 (N^w+N^{\bar{w}}) m^w \cbw_{TT} \\
m^{\rm B}_g &=& 
m^{\rm B} + \sum_w \Big( (N^w+N^{\bar{w}}) m^w \cbw_{TT}
+ 2 \al (N^w-N^{\bar{w}}) \afbw_T \Big).
\label{friedmasses}
\eea
Here B is either T or S,
$N^w$ and $N^{\bar{w}}$
denote the number of particles and antiparticles of type $w$,
respectively,
and $m^w$ is the mass of a particle of type $w$.
The IPM for electrons, protons, and neutrons,
is then defined by the three conditions
$\al \afbw_T = \fracn 13 m^w \cbw_{TT}$,
where $w$ ranges over $e$, $p$, $n$.

The defining conditions of the IPM
ensure that for a matter body B
the effective inertial and gravitational masses are equal.
That is,
$m^{\rm B}_i = m^{\rm B}_g$,
and hence no Lorentz-violating effects 
appear in gravitational tests to \pno3
using ordinary matter.
However,
for an antimatter test body T, 
$m^{\rm T}_i \neq m^{\rm T}_g$
within the IPM.
Thus, 
observable signals arise in comparisons 
of the gravitational responses of matter and antimatter
or between different types of antimatter.

Though the IPM was constructed
as a field-theoretic toy model
in which the indirect limitations on antimatter gravity
could be explored,
it is interesting to note the levels at which
anomalous effects in antimatter gravity experiments
might be seen.
The validity of the perturbative analysis
under which the present results are generated \cite{lvgap}
requires that the coefficients $\al \afbw_T = m^w \cbw_{TT}/3$
are perturbatively small relative to $m^w$. 
Values even as large as $0.5 m^w$
might be consistent with this restriction.
This would result in gravitational accelerations of hydrogen and antihydrogen 
that differ at the 50\% level.
However,
certain experiments with sensitivity to order $v^2$ effects
including the recent redshift analysis \cite{lvgap,mh}
and double-boost suppression terms,
if analyzed
in some flat spacetime tests,
could restrict the size considerably more.
Note however,
that these restrictions are considerably different
than the indirect experimental restrictions
usually considered,
and that they may be evaded in more realistic models
that could be created
via consideration of higher-order SME terms \cite{km-nr2}.
The following paragraphs
point out how
the IPM does not appear to limited by some of the usual arguments 
against anomalous antimatter gravity.

One issue that has been raised
is whether energy remains conserved
when antimatter has a different
gravitational response than does matter
\cite{pm}.
In the context of the present SME-based discussion,
the issue is moot because 
the energy-momentum tensor is explicitly conserved.
However,
as an illustration,
it is useful to consider a thought experiment 
in which a particle and an antiparticle
are lowered in a gravitational field,
converted to a photon pair,
raised to the original location,
and reconverted to the original particle-antiparticle pair.
Generically problems can arise
if the particle, antiparticle, and photons
each provide different contributions to the energy.
It is possible
to see explicitly how these issues are avoided.
In the analysis of Ref.\ \cite{lvgap},
the photons are conventional,
partly via an available coordinate choice.
The CPT-odd coefficient $\afbw_T$
does lead to different contributions to the energy,
but the two contributions cancel.
Contributions due to the CPT-even coefficient $\cbw_{TT}$ 
exist and combine during the lowering step;
however, the effect is offset by a 
$\cbw_{TT}$ contained in the definition of the conserved energy.

Neutral mesons
provide natural interferometers
mixing particle and antiparticle states.
Experiments studying such systems
could in principle also restrict
the gravitational response of antimatter \cite{mlg}. 
In the context of the SME in Minkowski spacetime,
neutral-meson systems have already been used to place tight constraints 
on certain differences of the coefficients $\afbw_\mu$
for $w$ ranging over quark flavors
\cite{mesons,akmesons}.
No dominant implications
arise from these constrains
for baryons,
which involve three valence quarks,
or for leptons.
Moreover
in meson tests,
valence $s$, $c$, or $b$ quarks
are involved,
which are largely irrelevant for protons and neutrons.
The same line of reasoning holds
for gravitational interactions.
Again the flavor dependence of Lorentz and CPT violation
implies that the IPM evades restrictions from meson systems. 

Finally,
consider the attempt to argue against
an anomalous gravitational response of antimatter
based on the large binding energy content of 
baryons, atoms, and bulk matter
\cite{lis2}.
A modern version of the argument 
relevant for the present discussion of antihydrogen
could proceed by noting
that the quarks in hydrogen contain only about 10\% of the mass,
with much of the remainder contained 
in the gluon and sea binding.
One might then concluding that 
since the binding forces are comparable 
for hydrogen and antihydrogen
the gravitational response of the two cannot differ 
by more than about 10\%.

Arguments such as the above
implicitly assume
that the gravitational response of a body is determined
by its mass and hence by binding energy.
In the IPM,
the coefficient $\afbw_T$,
leads to a correction to the gravitational force
that is independent of mass,
but can vary with flavor.
In fact,
the binding energies are largely conventional in the IPM,
and the Lorentz-violating modifications to the gravitational responses
are determined primarily by the flavor content
of the valence particles.
It is in principle even conceivable
that an anomalous gravity effect could be associated
purely with the positron,
as would occur in the IPM when $\afbe_T$ is the only nonzero coefficient.
More define statements
along the above lines would be possible
via careful consideration of radiative effects
involving $\afbw_T$, $\cbw_{TT}$,
and other SME coefficients for Lorentz violation 
\cite{ck,renorm},
perhaps imposing the IPM condition 
only after renormalization;
however,
the essential points illustrated with the IPM are:
the gravitational response of a body
can be independent of mass,
can vary with flavor,
and can differ between particles and antiparticles.

\section{Summary}

Searches for Lorentz and CPT violation
provide the opportunity to probe Planck-scale physics
with existing technology,
and a general field-theoretic framework for such investigations
is provided by the SME.
A large number of SME-based experiments
have been performed or proposed across many areas of physics,
including recent progress in gravitational physics.
The comparison of matter and antimatter,
including hydrogen and antihydrogen,
provides a means of conducting such investigations
using spectroscopic and gravitational experiments.
Moreover,
scenarios exist in which the antimatter studies
could detect Lorentz violation
that is otherwise challenging to observe.

\end{document}